\pdfoutput=1

\documentclass[11pt]{article}

\usepackage[]{acl}
\usepackage{times}
\usepackage{latexsym}
\usepackage{booktabs}
\usepackage{verbatim}
\usepackage{xcolor}
\usepackage[T1]{fontenc}

\usepackage[utf8]{inputenc}

\usepackage{microtype}
\definecolor{lightred}{rgb}{1.0, 0.8, 0.8}
\definecolor{lightblue}{rgb}{0.8, 0.9, 1.0}
\definecolor{lightgreen}{rgb}{0.8, 1.0, 0.8}
\definecolor{lightyellow}{rgb}{1.0, 1.0, 0.8}
\definecolor{lightpurple}{rgb}{0.9, 0.8, 1.0}
\definecolor{lightorange}{rgb}{1.0, 0.9, 0.8}
\usepackage{inconsolata}

\usepackage{amsmath, amssymb, graphicx, xcolor, multirow, multicol, comment, subcaption, algorithm2e, url, float, etoolbox, adjustbox, pgf, soul, geometry, colortbl, booktabs}

\usepackage{tcolorbox}

\title{\textit{Beyond Speech and More}: Investigating the Emergent Ability of Speech Foundation Models for Classifying Physiological Time-Series Signals}

\author{
  Orchid Chetia Phukan$^{1*}$, 
  Swarup Ranjan Behera$^{2}$\thanks{\footnotesize{First co-authors}},  
  Girish$^{1,3}$\thanks{\footnotesize{Second co-authors}}, 
  Mohd Mujtaba Akhtar$^{1\dagger}$\\
  \textbf{Arun Balaji Buduru}$^{1}$,
  \textbf{Rajesh Sharma}$^{1,4}$\\
  \textsuperscript{1}IIIT-Delhi, India, 
  \textsuperscript{2}Reliance Jio AICoE, India, 
  \textsuperscript{3}UPES, India,
  \textsuperscript{4}University of Tartu, Estonia\\ 
  \texttt{\textbf{Correspondence:} \textcolor{blue}{orchidp@iiitd.ac.in}}
}

\begin{document}
\maketitle


\begin{abstract}
Despite being trained exclusively on speech data, speech foundation models (SFMs) like Whisper have shown impressive performance in non-speech tasks such as audio classification. This is partly because speech shares some common traits with audio, enabling SFMs to transfer effectively. In this study, we push the boundaries by evaluating SFMs on a more challenging out-of-domain (OOD) task: classifying physiological time-series signals. We test two key hypotheses: first, that SFMs can generalize to physiological signals by capturing shared temporal patterns; second, that multilingual SFMs will outperform others due to their exposure to greater variability during pre-training, leading to more robust, generalized representations. Our experiments, conducted for stress recognition using ECG (Electrocardiogram), EMG (Electromyography), and EDA (Electrodermal Activity) signals, reveal that models trained on SFM-derived representations outperform those trained on raw physiological signals. Among all models, multilingual SFMs achieve the highest accuracy, supporting our hypothesis and demonstrating their OOD capabilities. This work positions SFMs as promising tools for new uncharted domains beyond speech.
    
\end{abstract}

\section{Introduction}

The dawn of large-scale pre-trained foundation models (FMs) has revolutionized machine learning across diverse modalities, including speech, audio, text, and image. These FMs, trained on vast datasets, possess exceptional generalization abilities, enabling them to perform a wide range of tasks both in-domain (ID) as well as out-of-domain (OOD). Large langauge models (LLMs) despite being pre-trained on text data have shown excellent OOD performance in predicting protein phase transition \cite{frank2024leveraging}, speech-based depression detection \cite{zhang2024llms}. Similarly, researchers have leveraged vision transformer pre-trained on visual data for speech emotion recognition \cite{akinpelu2024enhanced}. In audio\footnote{Here, audio is categorized as sounds that are non-speech} domain, speech foundation models (SFMs) primarily trained on large scale speech have exhibited cross-task generalization, handling tasks like audio classification \cite{gong23d_interspeech, ma-etal-2024-investigating}, bio-acoustics \cite{sheikh24_interspeech}, etc. These advancements underscore the emergent capabilities of FMs to extend beyond their original training domains, unlocking new avenues for exploration and application across disciplines.

In this study, we focus especially on investigating the ability of SFMs to perform a challenging OOD task: classification of physiological time-series signals. Previously, researchers in NLP and vision community have explored the usage of text-based FMs and vision-based FMs for physiological signals applications \cite{gao2024raw, 10.1145/3675094.3678494, yoon2024eyesgroundingmultimodallarge, phukan2024sonic}. However, no effort has yet been made to use SFMs to classify physiological signals.
We address the gap by evaluating the implicit capabilities of SFMs to classify physiological signals for stress recognition. We hypothesize that: (i) \textit{SFMs can generalize to classify physiological signals by leveraging the shared temporal patterns inherent in both speech and physiological time-series data} and (ii) \textit{Multilingual SFMs will perform better than other SFMs due to their large-scale pre-training on diverse languages equips them with robust, generalized representations that enhance their ability to capture complex temporal patterns, making them more effective.} \par

To test these hypotheses, we consider three different types of physiological signals, including ECG (Electrocardiogram), EMG (Electromyography), and EDA (Electrodermal Activity), for a better understanding of the OOD generalizability of the SFMs. We conduct experiments on the WESAD dataset, a benchmark for physiological stress recognition. We extract representations from the frozen SFMs and show that downstream models trained on SFM representations outperform models trained on raw physiological data. Amongst SFMs, we show that multilingual SFMs achieve the best performance. Our results validate the OOD capabilities of SFMs and underscore the potential of SFMs for usage to physiological signal applications, opening new possibilities. The main contributions are summarized as follows:

\begin{itemize}
\item We are, to the best of our knowledge, the first study to evaluate diverse state-of-the-art (SOTA) SFMs - WavLM, Wav2vec2, Unispeech-SAT, x-vector, HuBERT, MMS, XLS-R, and Whisper - for classifying physiological signals such as ECG, EMG, and EDA for stress recognition.
\item Our findings show that SFMs representations outperform models trained on raw physiological data. This mirrors findings in NLP, where LLM representations have shown competitive performance compared to raw data ~\cite{gao2024raw}.

\item We demonstrate that multilingual SFMs exhibit the best performance among all the SOTA SFMs. This superior performance is consistent across the three physiological signals (ECG, EMG, and EDA) under consideration.
\end{itemize}

\noindent For reproducibility, we will release the codes curated as part of this study after the double-blind review process.

\section{Related Work}
In this section, we briefly explore the application of SFMs for tasks beyond speech, followed by an examination of FMs from other domains utilized for physiological time-series signal tasks.

SFMs have shown exceptional performance in diverse tasks outside speech processing. \citet{sarkar23_interspeech} et al. investigated various SOTA SFMs such as WavLM, Wav2vec2, etc for classifying animal callers. \citet{cauzinille24_interspeech} et al. evaluated self-supervised SFMs for classifying gibbon's vocal signatures. \citet{turian2022hear} explored various SFMs representations for achieving SOTA performance across various acoustic tasks such as environmental sound classification, music tasks, and so on. However, the audio tasks, as well as the speech processing applications, fall under the broader domain of sounds. This leaves a gap in understanding the performance of SFMs in OOD tasks. In our study, we focus on a challenging OOD task: the classification of physiological signals with diverse applications ranging from affective computing to healthcare. 

Researchers from NLP and CV domains have explored the usage of their domain-specific FMs for classifying physiological signals. \citet{10.1145/3675094.3678494} exploited LLM for mental health assessment with EEG signals and \citet{ishaque2022detecting} leveraged various vision foundation models such as ResNet, VGG-16, etc for stress recognition from ECG signals. However, SFMs haven't been explored yet for the classification of physiological signals, and in this work, we take the first step towards this direction by classifying physiological signals for stress recognition.




\section{Speech Foundation Models}

In this section, we give an overview of the different SOTA SFMs considered in our study. \textbf{WavLM~\cite{chen2022wavlm}} is a SOTA SFM on SUPERB that integrates masked speech prediction and denoising. \textbf{Wav2vec2~\cite{baevski2020wav2vec}} is a self-supervised learning model that solves a contrastive task. \textbf{UniSpeech~\cite{Chen2021UniSpeechSAT}} is also a SOTA SFM on SUPERB, pre-trained in a speaker-aware format. \textbf{x-vector~\cite{snyder2018x}} is a time-delay neural network trained for SOTA speaker recognition. \textbf{HuBERT~\cite{hsu2021hubert}} learns from continuous speech without labeled data, refining clusters through a masked prediction loss. \textbf{MMS~\cite{pratap2024scaling}} expands speech technology to over 1,000 languages and is built upon Wav2vec2 architecture. \textbf{XLS-R~\cite{babu2021xlsr}} is a cross-lingual model trained on 128 languages and based on Wav2vec2 architecture. \textbf{Whisper~\cite{radford2022whisper}} is an encoder-decoder-based multilingual transformer-based model trained on 96 languages. \par

We use the frozen SFMs and extract representations from the last hidden layer of the SFMs by average pooling. We extract representations of 512 (x-vector, Whisper), 768 (WavLM, Wav2vec2, Unispeech-SAT, HuBERT), 1024 (WavLM-large), and 1280 (MMS, XLS-R) - dimension from the SFMs. For Whisper, we leveraged the representations from the encoder by removing the decoder. More detailed information and the links to the SFMs are given in Appendix~\ref{sec:appendix1}.

\begin{figure}[bt]
    \centering
    \includegraphics[width=1\linewidth]{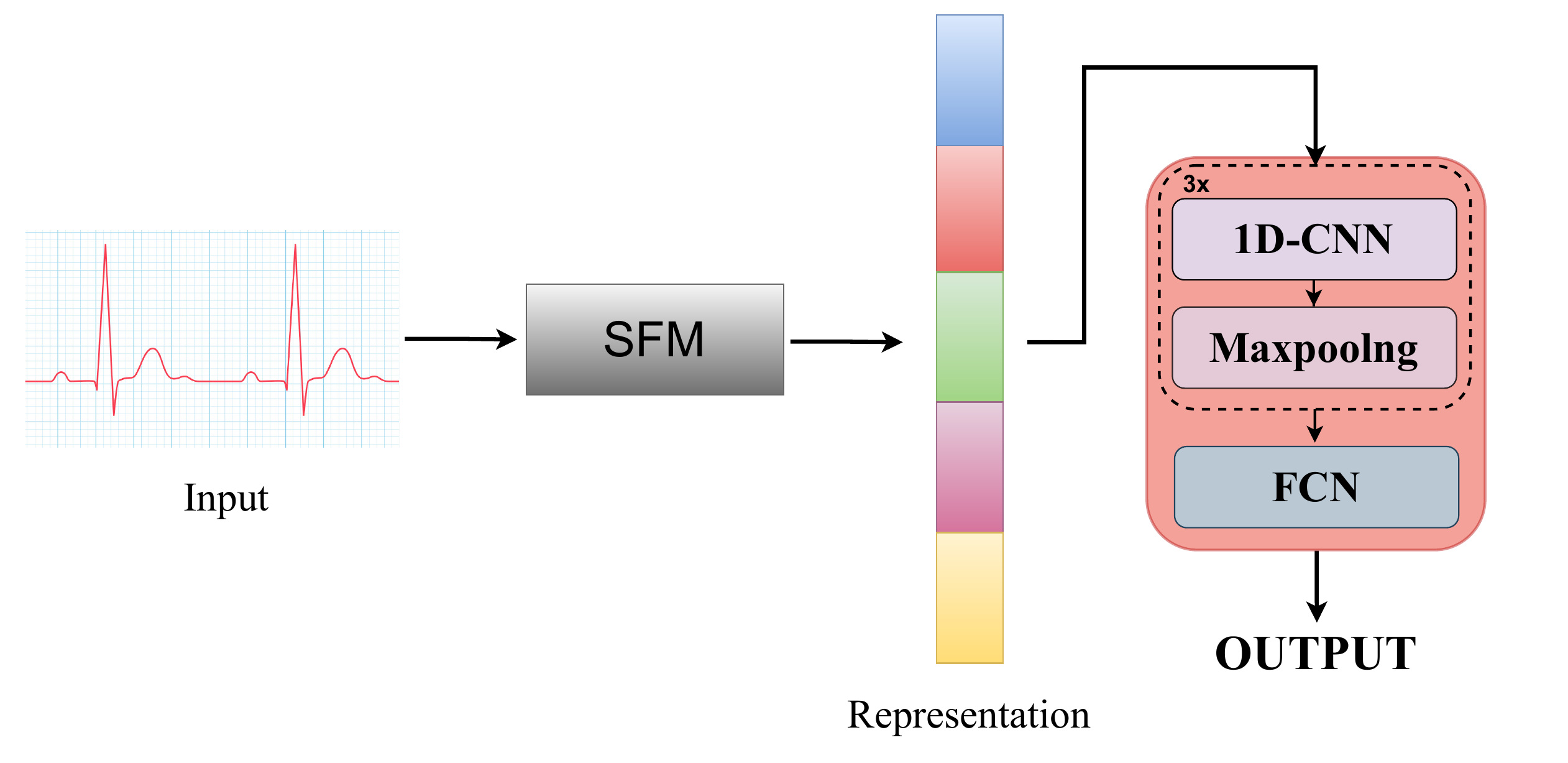}
    \caption{Modeling architecture.}
    \label{archi}
\end{figure}

\section{Experimental Setup}

\subsection{Benchmark Dataset} 
We use the WESAD (Wearable Stress and Affect Detection) dataset~\cite{schmidt2018introducing}. 
It consists of physiological signals from chest and wrist-worn sensors. Collected from 15 subjects at a 700 Hz sampling rate, the dataset supports both 2-class (stress vs. non-stress) and 3-class (baseline vs stress vs amusement) classification tasks. For our experiments, we leverage ECG, EMG, and EDA signals collected from chest-worn sensors. We make windows of 5 seconds with a shift of 2 seconds following \citet{singh2023stress}. Further, as the SFMs expect the input to be sampled 16 KHz, we resample the signals to 16 KHz. We then directly pass the signals to the SFMs for the representations to be extracted for downstream modeling. 

\subsection{Downstream Modeling}
We kept the downstream modeling simple to investigate the implicit behavior of SFMs for classifying physiological signals. We employ CNN and FCN (Fully Connected Network) as downstream networks and built models for both 2-class and 3-class classification. The CNN (Figure~\ref{archi}) model consists of two 1D convolutional layers with 16, 32, and 64 filters, each followed by batch normalization and max-pooling. Then, we flattened the output and passed it through an FCN with a dense layer with 128 neurons. We keep the number of neurons in the output layer depending on the classification task i.e 2-class or 3-class classification that outputs the class probabilities. For the FCN, we follow the same modeling as for FCN in the CNN model. We train the models for 50 epochs with a learning rate of 1e-3 and use Adam as the optimizer. To prevent overfitting, we use dropout and early stopping.

\begin{table}[bt]
\scriptsize
\centering
\begin{tabular}{l|c|c|c|c}
\toprule
                                 & \multicolumn{2}{c|}{\textbf{FCN}}            & \multicolumn{2}{c}{\textbf{CNN}}            \\
 \midrule
                                 & \textbf{A (\%)}     & \textbf{F1 (\%)}     & \textbf{A (\%)}      & \textbf{F1 (\%)}     \\
\midrule
\multicolumn{5}{c}{\textbf{2-class}} \\ 
\midrule
\textbf{ECG}                     & 58.71      & 58.68         &   75.42      & 64.18            \\
\textbf{EMG}                     &  72.17      & 54.42            &      72.90      & 62.12         \\
\textbf{EDA}                     & 39.38               & 36.74               & 55.26               & 55.20               \\
\midrule
\multicolumn{5}{c}{\textbf{3-class}} \\ 
\midrule
\textbf{ECG}                     &  44.26      & 36.18        &  57.78      & 44.30              \\
\textbf{EMG}                     & 40.32      & 35.51        & 58.05      & 45.72         \\
\textbf{EDA}                     & 33.25               & 21.02               & 46.23               & 33.04               \\
\bottomrule
\end{tabular}
\caption{Evaluation Scores: A and F1 represent accuracy and macro-average F1 scores, respectively, for both 2-class and 3-class classification. The scores are averages from a 5-fold evaluation. All scores are in \%.}
\label{tab:performance_scores1}
\end{table}

\begin{table*}[bt]
\scriptsize
\centering
\begin{tabular}{l|c|c|c|c|c|c|c|c|c|c|c|c c|c}
\toprule
            & \multicolumn{6}{c|}{\textbf{FCN}}                                     & \multicolumn{6}{c}{\textbf{CNN}}                                      \\
\midrule
            & \multicolumn{2}{c|}{\textbf{ECG}} & \multicolumn{2}{c|}{\textbf{EMG}} & \multicolumn{2}{c|}{\textbf{EDA}} & \multicolumn{2}{c|}{\textbf{ECG}} & \multicolumn{2}{c|}{\textbf{EMG}} & \multicolumn{2}{c}{\textbf{EDA}} \\
\midrule
\textbf{SFMs} & \textbf{A (\%)}    & \textbf{F1 (\%)}   & \textbf{A (\%)}    & \textbf{F1 (\%)}   & \textbf{A (\%)}    & \textbf{F1 (\%)}   & \textbf{A (\%)}    & \textbf{F1 (\%)}   & \textbf{A (\%)}    & \textbf{F1 (\%)}   & \textbf{A (\%)}    & \textbf{F1 (\%)}   \\
\midrule
\textbf{WavLM} & 81.60      & 73.42      & 78.05      & 69.99     & 64.26      & 43.45     & 81.90      & 76.39      & 78.66      & 71.36      & 70.29      & 63.70      \\
\textbf{WavLM-large}       & 85.00      & 80.70      & 73.78      & 60.89     & 67.23      & 42.02      & 89.78      & 88.18   & 76.69      &   70.97      & 69.69      & 55.76 \\
\textbf{Wav2vec2} & 80.45      & 75.59        & 75.17      & 66.99     & 68.17      & 49.99       & 84.21      & 81.47    & 77.24      &  72.23     & 71.45      & 61.67      \\
\textbf{Unispeech-SAT}         & 76.63      & 65.07      & 72.45      & 54.84       & 64.99      & 41.20     & 79.60      & 75.89     & 75.73      & 66.56     & 69.78      & 55.23     \\
\textbf{x-vector}     & 83.42      & 78.86     & 75.87     & 60.33      & 73.54    &     55.34       & 86.78       & 82.92     &  75.90     &  66.24     &74.72      & 63.28  \\
\textbf{HuBERT}      & 82.93      & 76.84      &  75.17     & 72.57      &   73.02     & 55.21     & 84.51      & 82.97      & 79.78      & 73.57     &  74.96      & 63.09    \\
\textbf{MMS}         & \cellcolor{lightpurple}\textbf{86.94}     & \cellcolor{lightpurple}\textbf{83.08}      & \cellcolor{lightblue}\textbf{80.18}      & \cellcolor{lightpurple}\textbf{72.82}     &   \cellcolor{lightblue}\textbf{75.08}      &  \cellcolor{lightblue}\textbf{64.54}    &  \cellcolor{lightpurple}\textbf{91.51}      & \cellcolor{lightpurple}\textbf{89.94}     & \cellcolor{lightblue}\textbf{81.39}      &   \cellcolor{lightblue}\textbf{77.46}     &   \cellcolor{lightgreen}\textbf{79.51}      & \cellcolor{lightgreen}\textbf{72.23}   \\
\textbf{XLS-R}        & \cellcolor{lightblue}\textbf{91.48}               & \cellcolor{lightblue}\textbf{89.21}      & \cellcolor{lightpurple}\textbf{79.63}      &  \cellcolor{lightblue}\textbf{73.40}      & \cellcolor{lightpurple}\textbf{74.33}      &  \cellcolor{lightpurple}\textbf{63.06}       & \cellcolor{lightblue}\textbf{91.69} & \cellcolor{lightblue}\textbf{90.22}     & \cellcolor{lightpurple}\textbf{80.21}      & \cellcolor{lightpurple}\textbf{76.83}      &  \cellcolor{lightblue}\textbf{78.33}      & \cellcolor{lightblue}\textbf{71.96}  \\
\textbf{Whisper}     & \cellcolor{lightgreen}\textbf{93.42}      & \cellcolor{lightgreen}\textbf{91.94}        & \cellcolor{lightgreen}\textbf{85.15}      & \cellcolor{lightgreen}\textbf{80.90}        & \cellcolor{lightgreen}\textbf{75.72}     &    \cellcolor{lightgreen}\textbf{67.57}     & \cellcolor{lightgreen}\textbf{97.85}      & \cellcolor{lightgreen}\textbf{97.43}     & \cellcolor{lightgreen}\textbf{87.00}      & \cellcolor{lightgreen}\textbf{83.92}    &  \cellcolor{lightpurple}\textbf{77.42}     &   \cellcolor{lightpurple}\textbf{68.14}  \\
\bottomrule
\end{tabular}
\caption{Evaluation Scores: A and F1 represent accuracy and macro-average F1 scores, respectively, for 2-class classification (stress vs. non-stress) with downstream models trained on different SFM representations. Scores are averages from a 5-fold evaluation. All scores are in \%.}
\label{tab:performance_scores2}
\end{table*}

\begin{table*}[bt]
\scriptsize
\centering
\begin{tabular}{l|c|c|c|c|c|c|c|c|c|c|c|c c|c}
\toprule
            & \multicolumn{6}{c|}{\textbf{FCN}}                                     & \multicolumn{6}{c}{\textbf{CNN}}                                      \\
\midrule
            & \multicolumn{2}{c|}{\textbf{ECG}} & \multicolumn{2}{c|}{\textbf{EMG}} & \multicolumn{2}{c|}{\textbf{EDA}} & \multicolumn{2}{c|}{\textbf{ECG}} & \multicolumn{2}{c|}{\textbf{EMG}} & \multicolumn{2}{c}{\textbf{EDA}} \\
\midrule
\textbf{SFMs} & \textbf{A (\%)}    & \textbf{F1 (\%)}   & \textbf{A (\%)}    & \textbf{F1 (\%)}   & \textbf{A (\%)}    & \textbf{F1 (\%)}   & \textbf{A (\%)}    & \textbf{F1 (\%)}   & \textbf{A (\%)}    & \textbf{F1 (\%)}   & \textbf{A (\%)}    & \textbf{F1 (\%)}   \\
\midrule
\textbf{WavLM}       & 52.89      & 39.08     & 44.59      & 39.03       & 49.86      &   27.36   & 67.02     & 49.27     & 58.93      &  47.86         & 53.14      & 38.15      \\
\textbf{WavLM-large} & 66.75      & 48.12     & 56.05      &  38.14     & 53.11      & 35.31       & 69.41      & 50.20      & 64.35               & 50.35     & 56.47      &  46.32  \\
\textbf{Wav2vec2} & 65.91      & 47.56       & 62.32      & 44.17      &  53.44      & 36.26      & 70.63      & 63.43     & 63.69      & 54.28      & 57.53     & 47.44       \\
\textbf{Unispeech-SAT}       &  63.23      & 44.21     & 51.53      & 36.72       & 46.14      & 24.55      & 64.47      & 56.31      & 66.18 & 48.05    & 53.11      &  39.37     \\
\textbf{x-vector}     &  69.11      & 56.02    & 59.35      & 40.08     & 45.32       & 43.82       & 76.72      & 67.60     & 61.41      & 53.56      & 59.62      & 41.38   \\
\textbf{HuBERT}      & 66.69      & 49.63       & 61.44      & 46.91     & 56.08      &   42.06      & 67.78      &  63.35     & 65.99      & 53.13       & 59.84      &43.38     \\
\textbf{MMS}         &   \cellcolor{lightpurple}\textbf{72.42}      & \cellcolor{lightpurple}\textbf{60.61}    & \cellcolor{lightpurple}\textbf{64.02}      & \cellcolor{lightpurple}\textbf{47.35}      & \cellcolor{lightpurple}\textbf{59.68}      & \cellcolor{lightpurple}\textbf{44.57}      & \cellcolor{lightblue}\textbf{78.27}      & \cellcolor{lightblue}\textbf{72.03}      & \cellcolor{lightblue}\textbf{68.29}               & \cellcolor{lightpurple}\textbf{56.32}     & \cellcolor{lightblue}\textbf{62.93}      &   \cellcolor{lightpurple}\textbf{49.06}     \\
\textbf{XLS-R}        & \cellcolor{lightblue}\textbf{74.96}  &  \cellcolor{lightblue}\textbf{66.29}    & \cellcolor{lightblue}\textbf{66.02}      &  \cellcolor{lightblue}\textbf{47.39}      &  \cellcolor{lightgreen}\textbf{60.69} & \cellcolor{lightgreen}\textbf{45.38}      & \cellcolor{lightpurple}\textbf{77.81}               &   \cellcolor{lightpurple}\textbf{69.06}     & \cellcolor{lightpurple}\textbf{66.87}      & \cellcolor{lightblue}\textbf{58.18}     &  \cellcolor{lightpurple}\textbf{61.72}      & \cellcolor{lightblue}\textbf{50.27}      \\
\textbf{Whisper}     & \cellcolor{lightgreen}\textbf{81.45}      & \cellcolor{lightgreen}\textbf{72.43}      & \cellcolor{lightgreen}\textbf{74.84}       & \cellcolor{lightgreen}\textbf{67.10}     & \cellcolor{lightblue}\textbf{60.08}       & \cellcolor{lightblue}\textbf{44.89}      & \cellcolor{lightgreen}\textbf{87.27}      & \cellcolor{lightgreen}\textbf{85.40}      & \cellcolor{lightgreen}\textbf{78.60}      & \cellcolor{lightgreen}\textbf{74.69}    & \cellcolor{lightgreen}\textbf{63.56}    & \cellcolor{lightgreen}\textbf{52.38}      \\
\bottomrule
\end{tabular}
\caption{Evaluation Scores: A and F1 represent accuracy and macro-average F1 scores, respectively, for 3-class classification (baseline vs. stress vs. amusement) with downstream models trained on different SFM representations. The scores are averages from a 5-fold evaluation. All scores are in \%.}
\label{tab:performance_scores3}
\end{table*}

\subsection{Experimental Results}
Table \ref{tab:performance_scores1} presents the performance of models trained on raw physiological data. Previous research in phsyiological signals mostly followed building models on raw signals \cite{singh2023stress, tanwar2024attention}, so we consider it as baseline in our study. In both 2-class classification (stress vs. non-stress) and 3-class classifcation (amusement, stress, base-
line), the CNN models achieve the best scores across ECG, EMG, and EMG compared to FCN.


Table \ref{tab:performance_scores2} and \ref{tab:performance_scores3} present the evaluation scores of a 2-class classification task (stress vs. non-stress) and 3-class classification task (baseline vs stress vs amusement), respectively. The scores demonstrate clear dominance of downstream models trained on SFM representations compared to models trained on raw data (See Table \ref{tab:performance_scores1}). This proves our first hypothesis that \textit{SFMs will generalize for classifying physiological signals by utilizing the common temporal patterns in native to both speech and physiological time-series data.} This performance of the SFMs is observed across all the considered SFMs, all the physiological signals, and also for both the classification tasks. \par

We considered two versions of WavLM - base and large, referred to as WavLM and WavLM-large. We chose WavLM-large for its SOTA performance in SUPERB \cite{yang21c_interspeech}, which assesses SFM representations for speech-processing tasks.
We find that WavLM-large outperforms WavLM, a difference likely attributable to the sizes of the SFMs: the base version has 94.70 million parameters, while the large version boasts 316.62 million.
However, when compared to other SFMs, it cannot be definitively stated that larger models always provide better representations. For instance, the base versions of Unispeech-SAT and HuBERT, which are significantly smaller than WavLM-large, demonstrate superior performance in some cases and are comparable in others.

Another interesting observation from the experimental results is the performance of x-vector, which, despite having only approximately 4.2 million parameters, outperforms many larger SFM counterparts. This highlights that SFMs trained to capture speaker characteristics can provide robust cross-domain representations.

Among all the SFMs in both Table \ref{tab:performance_scores2} and Table \ref{tab:performance_scores3}, we observe that multilingual SFMs outperform other models, including monolingual SFMs like Unispeech-SAT, WavLM, WavLM-large, and the speaker recognition PTM x-vector.
This validates our second hypothesis that \textit{multilingual SFMs will outperform other SFMs due to their large-scale pre-training across a wide variety of languages, making them adapt more effectively to the nuances of physiological signals and will provide more generalized cross-domain representations. Thus improving their ability to capture complex temporal patterns and enhancing their overall effectiveness.} 
This is further supported by the visualization of raw representations from the last hidden states of the SFMs using t-SNE plots in Appendix~\ref{sec:appendix1}, Figure \ref{fig:tsne}, where we observe better clustering across different classes in multilingual SFMs compared to others. However, among the multilingual SFMs, performance varies, with one model leading in some instances and another in others.

\vspace{-0.3cm}

\section{Conclusion}
In this study, we demonstrate for the first time that SFMs, despite being trained solely on speech data, exhibit notable effectiveness in the challenging OOD task of classifying physiological time-series signals for stress recognition. Our experiments on the WESAD dataset, using various SOTA SFMs, reveal that these models outperform those trained on raw physiological signals. Furthermore, we find that multilingual SFMs excel compared to counterparts, benefiting from multilingual pre-training that enhances their robustness and cross-domain generalization. Our findings open avenues for future research on the OOD generalization of SFMs.

\section{Limitations}
In our study, we evaluated SFMs using only two downstream networks. However, prior research in speech processing indicates that SFM performance varies across different architectures \cite{zaiem23b_interspeech}. In future work, we aim to explore a wider range of downstream models. Additionally, we focused solely on representations from the final layer of the SFMs, despite evidence suggesting that performance can fluctuate based on the extracted features from different layers \cite{kodali23_interspeech, pasad2021layer}. We will investigate this dimension in our future studies as well.

\section{Ethics Statement}
In our study, we utilize an openly accessible dataset that is freely available for research purposes. The dataset is fully anonymized, ensuring that no sensitive user details are associated with any of the data points. The dataset has been curated to meet ethical standards, and no additional processing was necessary to remove private user information. 

\bibliography{custom}

\appendix

\section{Appendix}
\label{sec:appendix}

\subsection{Speech Foundation Models} \label{sec:appendix1}


\noindent \textbf{Wav2Vec2}\footnote{\href{https://huggingface.co/facebook/wav2vec2-base}{https://huggingface.co/facebook/wav2vec2-base}}: Base version of 95.04 million
parameters that consists of 12 transformer encoder blocks.

\noindent \textbf{HuBERT}\footnote{\url{https://huggingface.co/facebook/hubert-base-ls960}}: Base version of 95 million parameters.

\noindent \textbf{WavLM}: We use base\footnote{\url{https://huggingface.co/microsoft/wavlm-base}} and large\footnote{\url{https://huggingface.co/microsoft/wavlm-large}} versions.

\noindent \textbf{UniSpeech-SAT}\footnote{\url{https://huggingface.co/microsoft/unispeech-sat-base}}: We utilized base version of 94.68 million parameters.

\noindent \textbf{x-vector}\footnote{\href{https://huggingface.co/speechbrain/spkrec-xvect-voxceleb}{https://huggingface.co/speechbrain/spkrec-xvect-voxceleb}}: It shows SOTA performance compared to previous SOTA speaker recognition system. It has approx 4.2 million parameters.

\noindent \textbf{MMS}\footnote{\url{https://huggingface.co/facebook/mms-1b}}: We make use of MMS 1 billion parameters version.

\noindent \textbf{XLS-R}\footnote{\url{https://huggingface.co/facebook/wav2vec2-xls-r-1b}}: We use XLS-R 1 billion parameters version.

\noindent \textbf{Whisper}\footnote{\url{https://huggingface.co/openai/whisper-base}}: We have selected the Whisper-base version of 74 million parameters for our experiments.


\begin{figure*}[bt]
    \centering
    \begin{minipage}{0.302\textwidth}
        \centering
        \includegraphics[width=\textwidth]{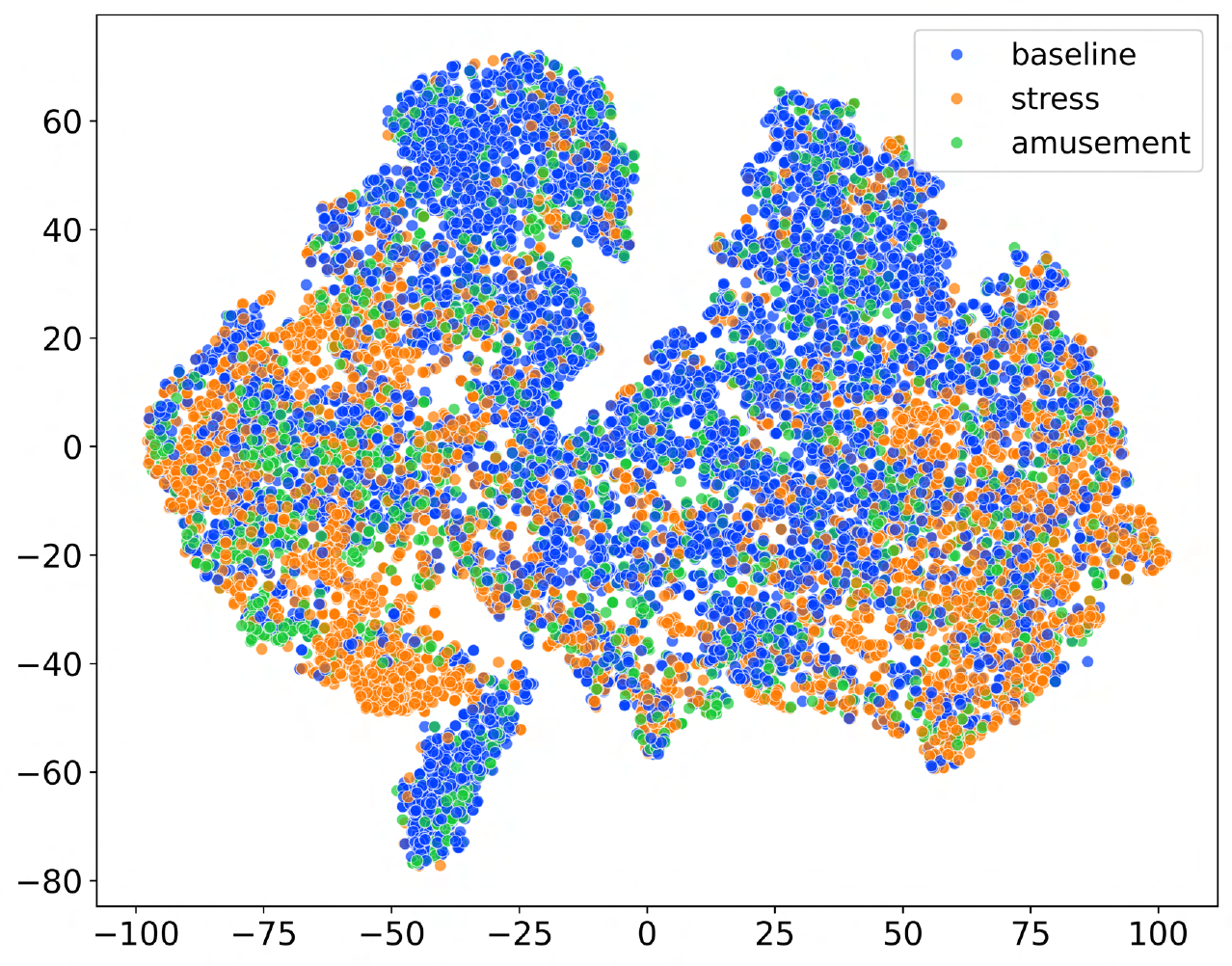}
        \caption*{(a) Wav2vec2 (EMG)}
    \end{minipage}\hfill
    \begin{minipage}{0.302\textwidth}
        \centering
        \includegraphics[width=\textwidth]{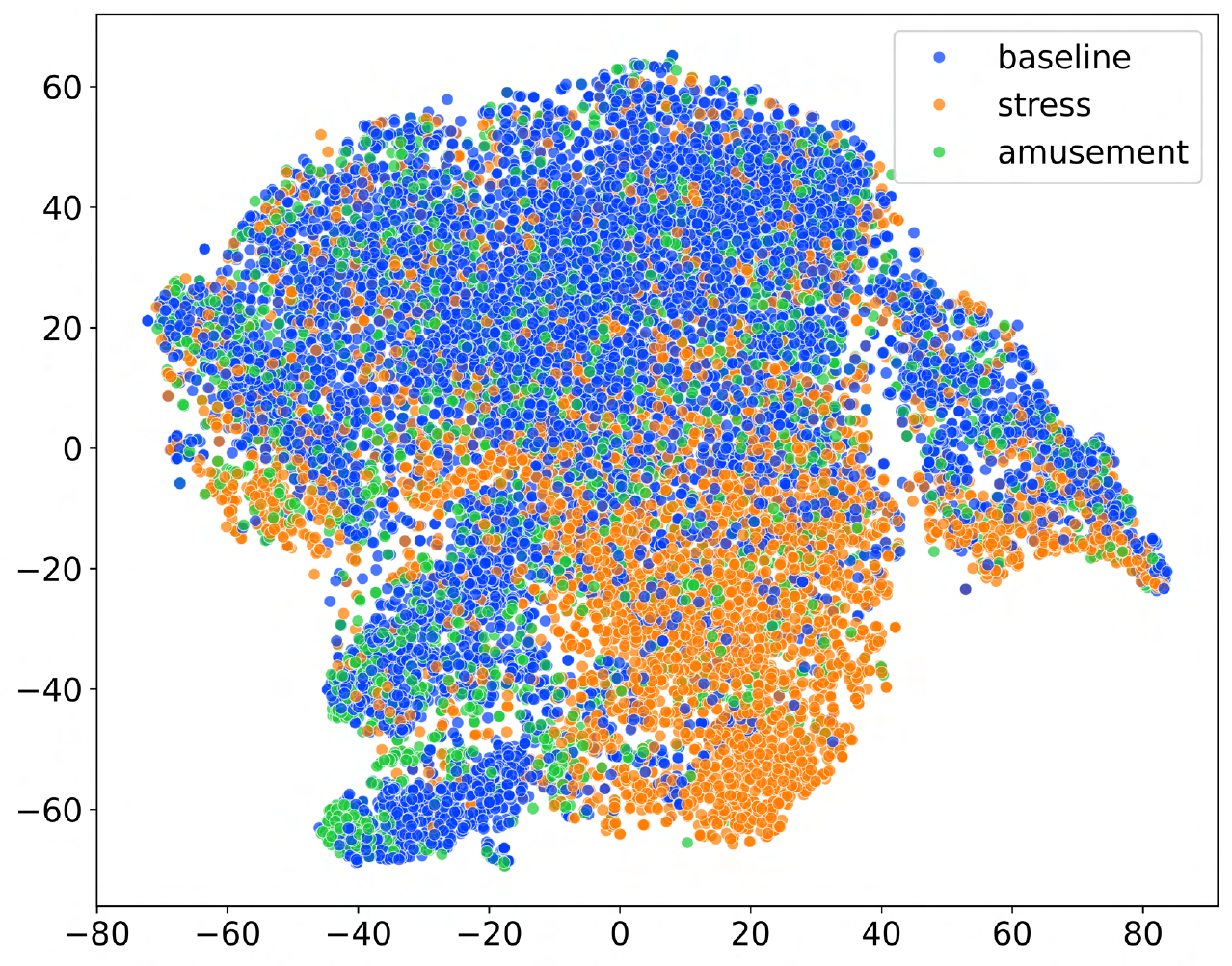}
        \caption*{(b) XLS-R (EMG)}
    \end{minipage}\hfill
    \begin{minipage}{0.302\textwidth}
        \centering
        \includegraphics[width=\textwidth]{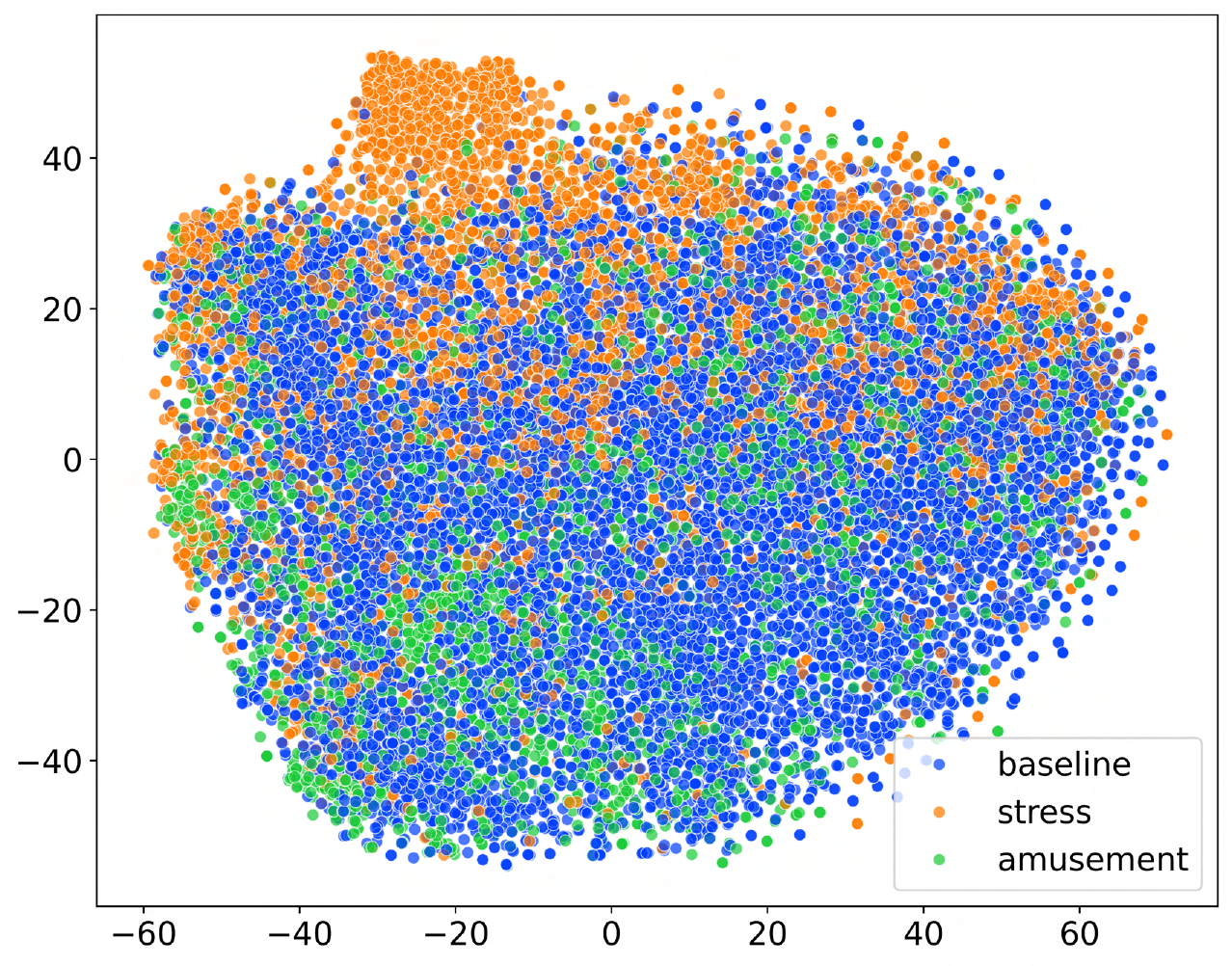}
        \caption*{(c) Whisper (EMG)}
    \end{minipage}\\[10pt] 
    \begin{minipage}{0.302\textwidth}
        \centering
        \includegraphics[width=\textwidth]{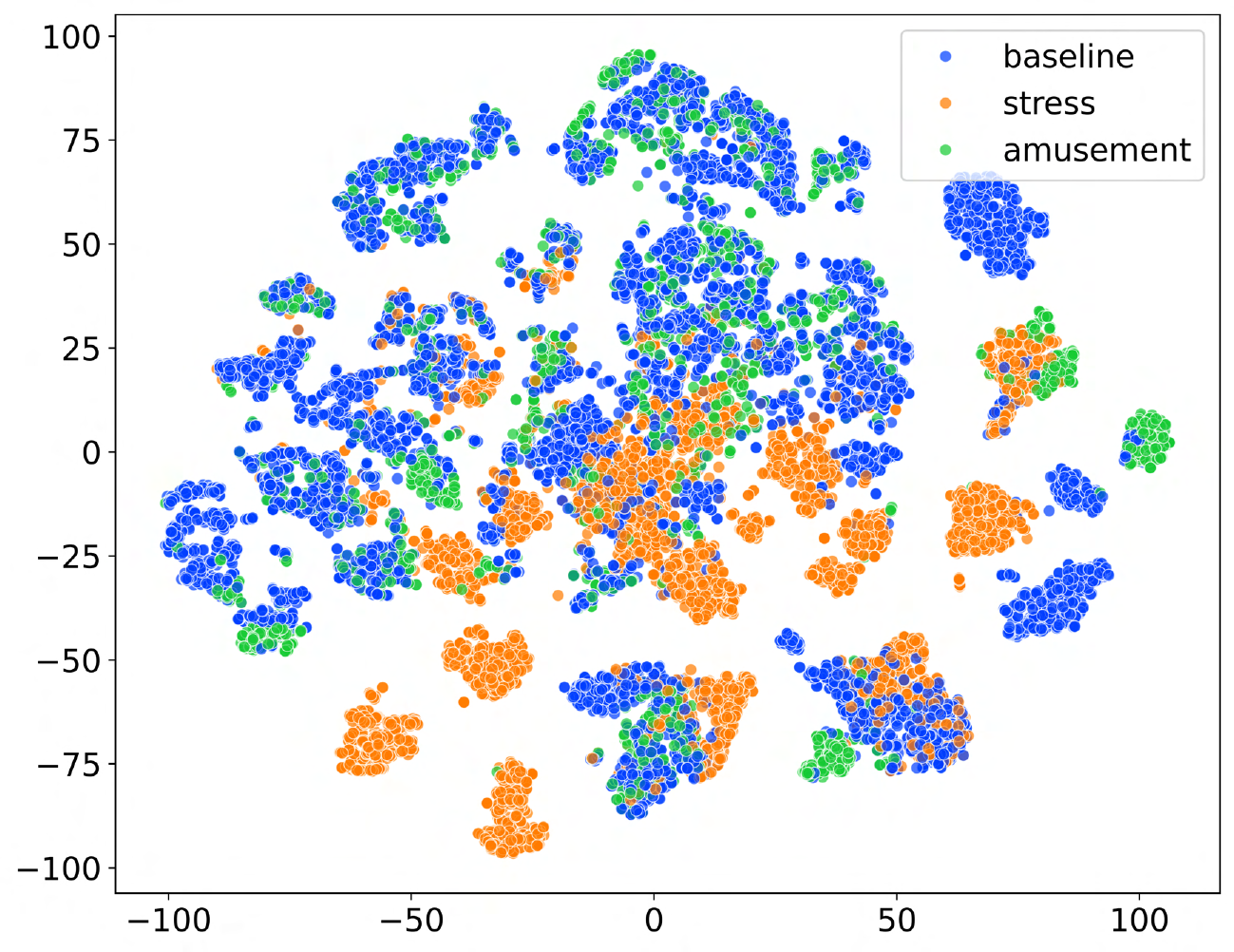}
        \caption*{(d) Whisper (ECG)}
    \end{minipage}\hfill
    \begin{minipage}{0.302\textwidth}
        \centering
        \includegraphics[width=\textwidth]{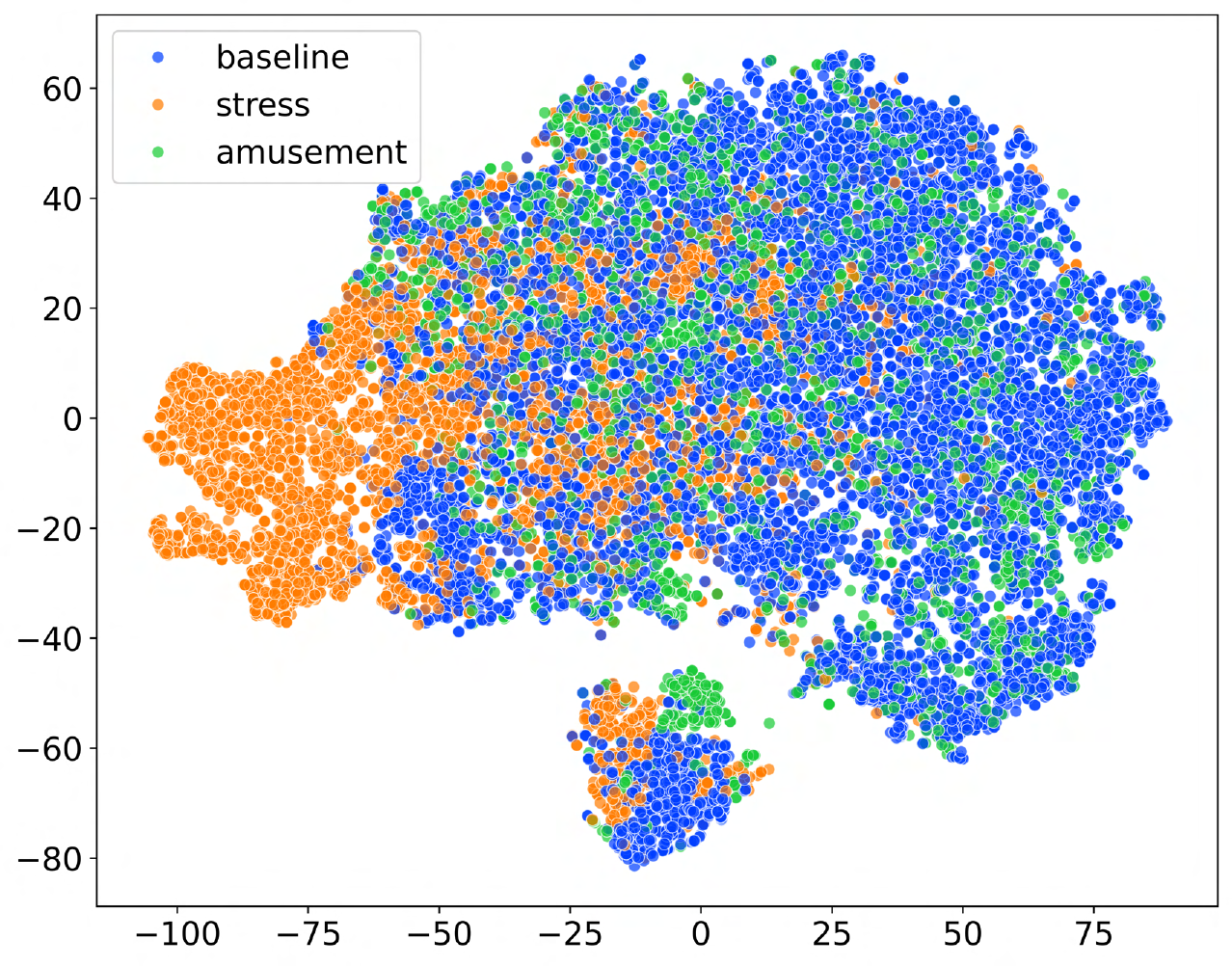}
        \caption*{(e) XLS-R (ECG)}
    \end{minipage}\hfill
    \begin{minipage}{0.302\textwidth}
        \centering
        \includegraphics[width=\textwidth]{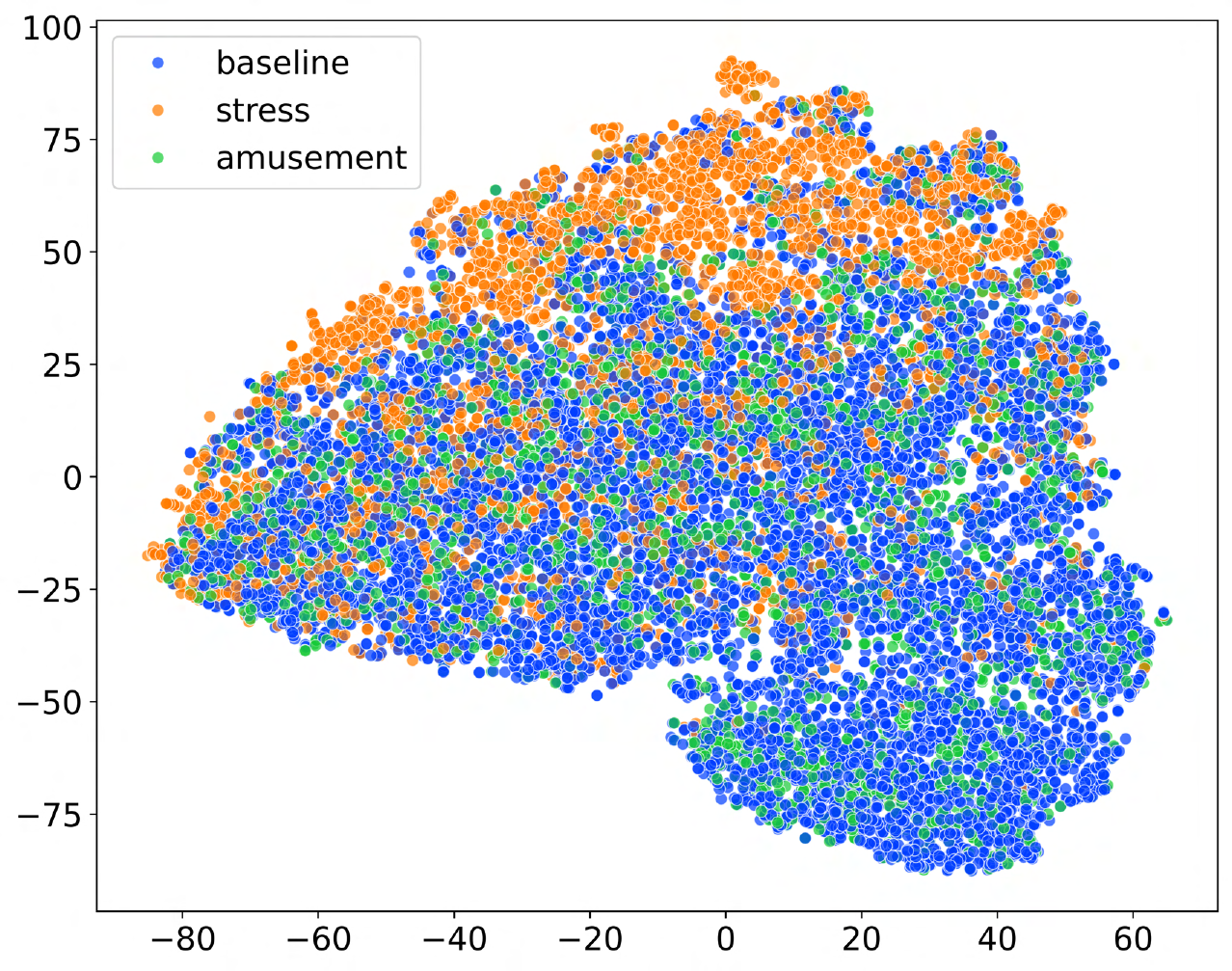} 
        \caption*{(f) MMS (ECG)}
    \end{minipage}\\[10pt] 
    \begin{minipage}{0.302\textwidth}
        \centering
        \includegraphics[width=\textwidth]{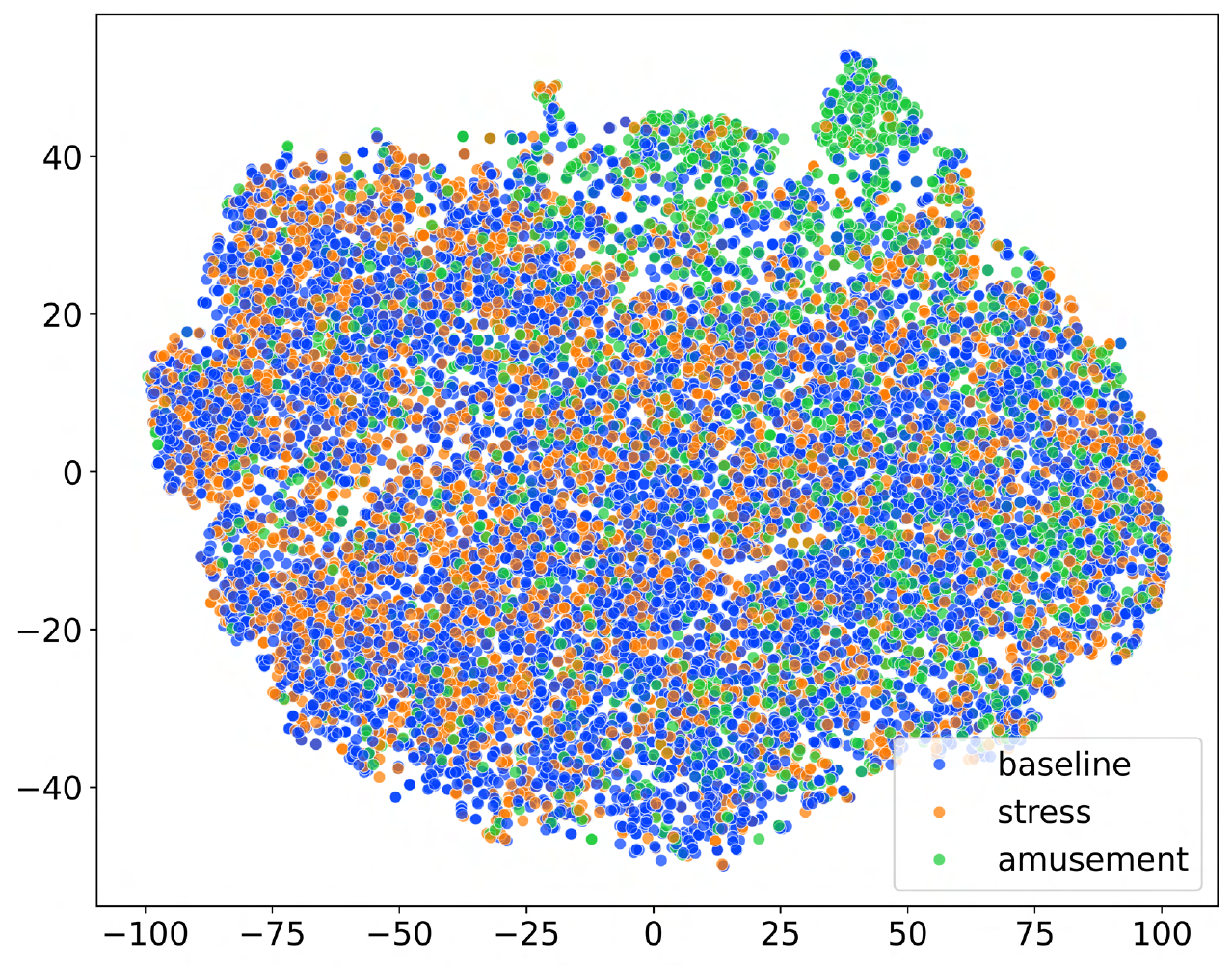}
        \caption*{(g) WavLM (EDA) }
    \end{minipage}\hfill
    \begin{minipage}{0.302\textwidth}
        \centering
        \includegraphics[width=\textwidth]{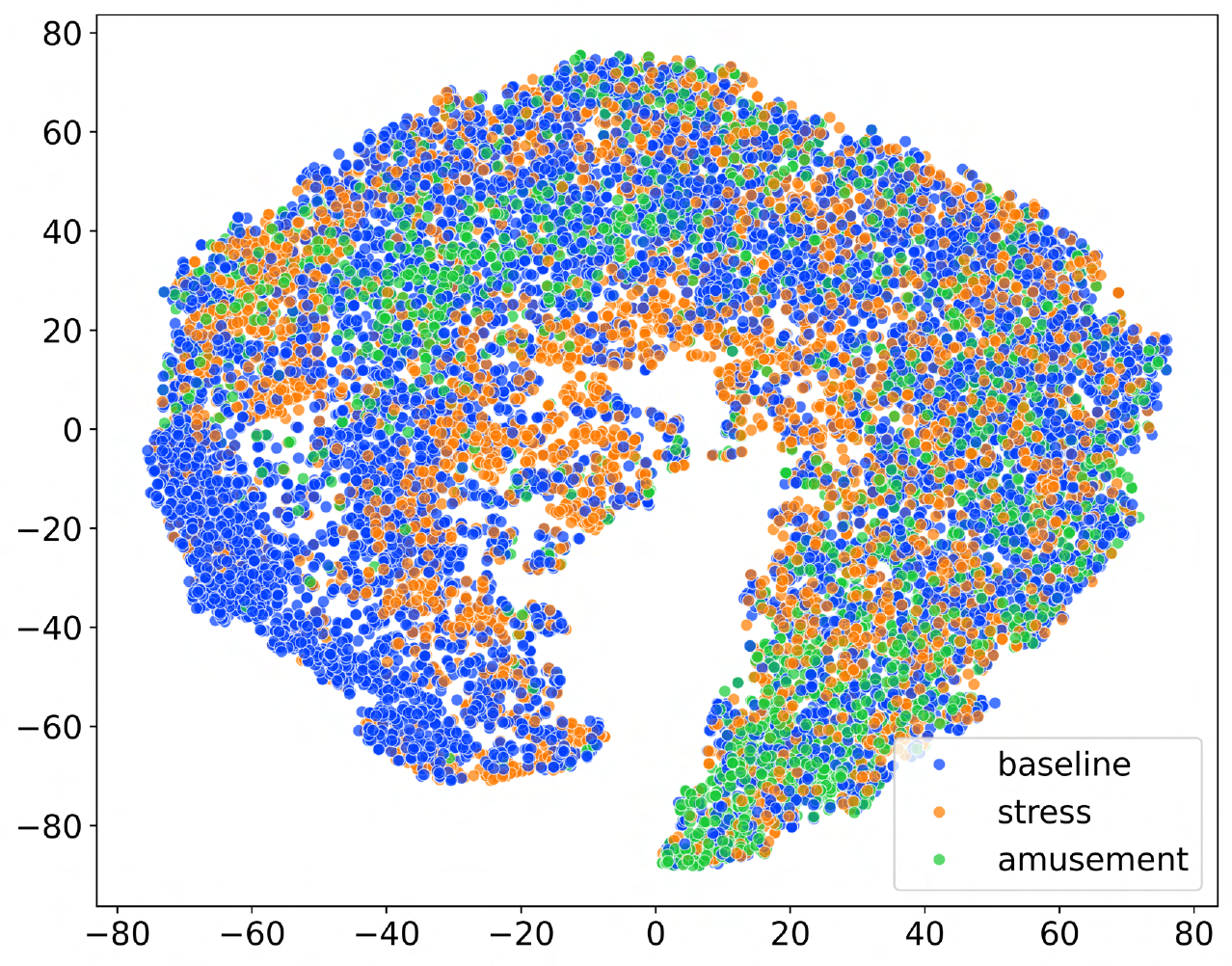} 
        \caption*{(h) Wav2vec2 (EDA)}
    \end{minipage}\hfill
    \begin{minipage}{0.302\textwidth}
        \centering
        \includegraphics[width=\textwidth]{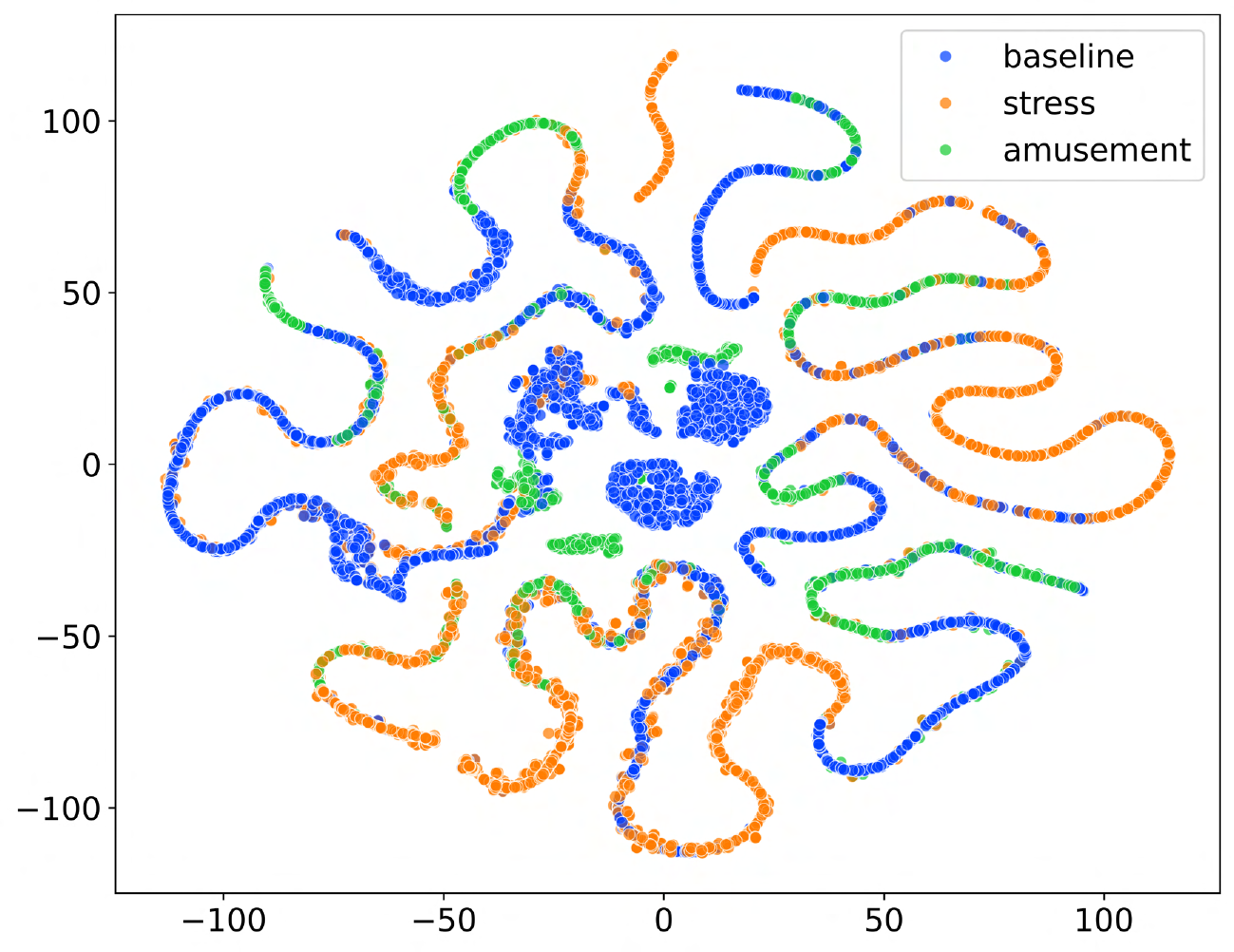} 
        \caption*{(i) Whisper (EDA)}
    \end{minipage}
    \caption{t-SNE Plots.}
    \label{fig:tsne}
\end{figure*}

\end{document}